\documentstyle[epsf,12pt]{article}
\input {epsf}
\makeatletter
\def\Lm{\Lambda}
\newcommand{\beq}{\begin{equation}}
\newcommand{\eeq}{\end{equation}}
\newcommand{\beqy}{\begin{eqnarray}}
\newcommand{\eeqy}{\end{eqnarray}}

\def\cW{{\cal W}}
\def\cWbar{\overline{{\cal W}}}

\def\wbar{\overline{w}}
\def\Wbar{\overline{W}}
\def\Abar{\bar{A}}
\def\Bbar{\bar{B}}
\def\Cbar{\bar{C}}

\catcode`\@=11
\def\numberbysection{\@addtoreset{equation}{section}
\def\theequation{\arabic{section}.\arabic{equation}}}
\def\appendix{\setcounter{section}{0}
        \def\thesection{Appendix \Alph{section}}
        \def\theequation{\Alph{section}.\arabic{equation}}}

\begin{document}
\begin{flushleft}
{\it  Yukawa Institute Kyoto}
\end{flushleft}
\begin{flushright}
 YITP-97-32\\
 hep-th/9706038\\
 June 1997
\end{flushright}
\renewcommand{\thefootnote}{\fnsymbol{footnote}}
\vspace{0.5in}
\begin{center}\Large{\bf 
Interaction of boundaries with heterogeneous matter states in
matrix models
}\\
\vspace{1cm}
\normalsize\ Masahiro Anazawa\footnote[1]
{Supported by JSPS. E-mail address: anazawa@yukawa.kyoto-u.ac.jp .},
Atushi Ishikawa\footnote[2]{
Supported by JSPS. E-mail address: ishikawa@yukawa.kyoto-u.ac.jp .}
$\quad$and$\quad$ Hirokazu Tanaka\footnote[3]{E-mail address:
hirokazu@yukawa.kyoto-u.ac.jp .}
\vspace{0.5in}

        Yukawa Institute for Theoretical Physics \\
        Kyoto University, Kyoto 606-01, Japan\\
\vspace{0.1in}
\end{center}
\renewcommand{\thefootnote}{\arabic{footnote}}
\setcounter{footnote}{0}
\vspace{0.4in}
\begin{abstract}
We study disk amplitudes whose boundary conditions 
on matter configurations are not restricted to homogeneous ones. 
They are examined in the two-matrix model
as well as in the three-matrix model for the case of the tricritical Ising
model. 
Comparing these amplitudes,
we demonstrate relations between degrees of freedom of matter
states in the two models.
We also show that they have a simple geometrical interpretation in terms of
 interactions of the boundaries. 
It plays an important role that two parts of a boundary with
different matter states stick each other.
We also find two closed sets of 
Schwinger-Dyson equations which determine disk amplitudes 
in the three-matrix model.
\end{abstract}

\vspace{0.2in}
\begin{flushleft}
\end{flushleft}

\newpage

\baselineskip 17pt

\numberbysection
\section{Introduction}
\label{introduction}

 It is well known that $(m,m+1)$ unitary conformal model
can be constructed microscopically as a lattice statistical
model\cite{Pasq}. 
At each site
this statistical model has
local degrees of freedom  labeled by 
the points of $A_{m-1}$ Dynkin diagram.
The $(m,m+1)$ model coupled to 2d gravity can be described 
by matrix model.  
The $(m,m+1)$ model has $(m-1)$ 
microscopic degrees of freedom and the $(m-1)$-matrix chain
model\cite{HI}
naturally corresponds to the $(m,m+1)$ model coupled to gravity.  
On the other hand
the two-matrix model\cite{KBK,Douglas,Tada,DKK}
can also describe this system near an appropriate
critical point though there are only two matrices.
In this paper we address the correspondence between the degrees of
freedom described by the matrices in the $(m-1)$-matrix chain model
and those in the two-matrix model\cite{AIT}.

 As a non-trivial simplest case, we study  
the $(4,5)$ tricritical Ising model coupled to 2d gravity.
As actions 
of the two- and the three-matrix models
which realize this system, we take 
\begin{eqnarray}
S(A,B) &=& \frac{N}{\Lambda} 
                  tr \left\{ U(A) + U(B) - AB \right\},
\label{2-matrix model Action}\\
S(A,B,C) & = & \frac{N}{\Lambda} 
                   tr \left\{ U_1(A) + U_2(B) + U_1(C) - A B - B C
                      \right\}.
\label{3-matrix model Action}
\end{eqnarray}

An amplitude of loops which have homogeneous matter configuration
corresponds to an expectation value of $tr( A^n)$ and so on.
The integration of the matrices
is to be separated into one  over eigenvalues
and one over angular variables.
We can integrate the angular variables first
and reduce the original integral to that in terms of the eigenvalues.  
Then through the  orthogonal polynomial method
we can show that the loop composed of
the matrix $A(B)$ in the two-matrix model corresponds to
that composed of the matrix $A(C)$ in the three-matrix model.
On the other hand, an amplitude of loops
with heterogeneous matter configurations
corresponds to an expectation value of 
$tr (A^n B^k \cdots$) and so on.
We cannot integrate the angular variables first in this case.
Then the argument in the case of homogeneous loops can not be applied,
and the correspondence between  the matrices in the two models is 
not so trivial in this case.
For simplicity we restrict our attention to disk amplitudes.
One of the purposes of the paper is to calculate disk amplitudes
with heterogeneous loops in the two models using Schwinger-Dyson 
technique\cite{GN,Staudacher,IIKMNS,SY}, and to study the correspondence between the matrices 
in the two- and three-matrix models. 

Studying these disk amplitudes is also a very interesting problem by
itself.  
We are forced to study heterogeneous loop 
when merging different homogeneous ones into a single one.
The amplitudes involving heterogeneous loops
have not been studied deeply\footnote{
In the case of Ising model these amplitudes were
 studied in \cite{Staudacher,SY}.}. 

We will obtain an interesting geometrical
picture on the amplitudes involving heterogeneous loops in sect.4.
If two parts of a loop 
have different matter states, these are forced to stick each
other and the original heterogeneous loop reduces to
homogeneous ones.
We obtain this picture through the study of the disk amplitudes.
We believe that this picture can  naturally be extended to 
cases with many loops or cases with many handles
because the sticking of boundaries of loops is a local phenomenon 
and does not depend on the global nature of surfaces.

The paper is organized as follows.
In sect.2 we compute  heterogeneous disk amplitudes whose loops are composed
of two arcs using the Schwinger-Dyson technique in the two-matrix model.
In sect.3 we compute  similar amplitudes in the three-matrix model.
In the process of our calculation we find two closed sets of Schwinger-Dyson
equations which are composed of seven and ten equations respectively.
The success of our calculation is due to these findings.
In sect.4 we discuss the correspondence between the matrices
in the two- and the three-matrix models, and
provide a geometrical picture on these amplitudes.
Sect.5 includes a summary and discussion.


\section{Two-matrix model case}
\label{2-matrix}

As a critical 
potential which realizes the $(4,5)$ model in the two-matrix model, 
we take
\begin{eqnarray}
U(\phi) &=& 8{\phi}+2{\phi}^2+\frac{8}{3}{\phi}^3+\frac{1}{4}{\phi}^4,
\label{2-matrix potential}
\end{eqnarray}
which can be determined by the method of orthogonal 
polynomial \cite{DKK} (see appendix). 
In this section, we would like to calculate the disk amplitude
\beq
W_{AB}(p,q)=
\frac{\Lm}{N} \bigl\langle tr\frac 1{p-A} \frac 1{q-B} \bigr\rangle
= \sum_{n,m=0}^{\infty} 
\frac{\Lm}{N}\left\langle tr(A^n B^m)\right\rangle
p^{-n-1}q^{-m-1}
\;,
\label{disk2AB}\eeq
and its continuum universal part
$w_{A B}(\zeta_A, \zeta_B, t)$ in the large $N$ limit
by means of the Schwinger-Dyson technique.
The boundary is consisted of two
parts which have different matter states.

Let us work out the disk amplitude
\begin{eqnarray}
W^{(B^m)}_{A}(p)=
\sum_{n=0}^{\infty} \frac{\Lm}{N}\left\langle tr(A^n B^m)\right\rangle
p^{-n-1},
\label{2-matrix resolv}
\end{eqnarray}
which will be necessary for the calculation of the amplitude (\ref{disk2AB}).
Consider the Schwinger-Dyson equations
\begin{eqnarray}
0 & = & \sum_a \int [dAdB] \frac{\partial}{\partial A^a} 
        \left\{
              tr \left( A^{n} t^a B^{k} \right) e^{-S(A,B)} 
        \right\},
\nonumber \\
0 & = & \sum_a \int [dAdB] \frac{\partial}{\partial B^a}
        \left\{ 
              tr \left( A^{n} t^a \right) e^{-S(A,B)}
        \right\},
\label{2-matrix S-D eq}
\end{eqnarray}
where 
we decomposed the matrices as 
$
A=\sum_{a=1}^{N^2}A^a t^a
$ etc. by introducing the basis $\{t^a\}$ of the hermitian matrix.
Using a notation 
$[A^n B^k] = \frac{\Lambda}{N}\langle tr (A^n B^k) \rangle$,
we can rewrite eqs.(\ref{2-matrix S-D eq}) as
\begin{eqnarray}
0 & = & \sum_{l=0}^{n-1} [A^l] [A^{n-1-l} B^k]
        - 8[A^{n} B^k] - 4[A^{n+1} B^k] - 8[A^{n+2} B^k]
        -  [A^{n+3} B^k] +  [A^{n} B^{k+1}],
\nonumber \\
0 & = & 8[A^n] + 4[A^n B] + 8[A^n B^2] + [A^n B^3] - [A^{n+1}],
\label{2-matrix S-D eq2}
\end{eqnarray}
in the large $N$ limit. 
It is convenient to use the resolvent representation
eq.(\ref{2-matrix resolv}) and we obtain
\begin{eqnarray}
0 & = & \left\{ W_A(p)-x(p) \right\} W^{(B^k)}_A(p) 
        + W^{(B^{k+1})}_A(p)+a^{(B^k)}(p),
\label{2-matrix S-D 1}
\\ 
0 & = & (8-p) W_A(p) + 4W^{(B)}_A(p) + 8W^{(B^2)}_A(p)
            + W^{(B^3)}_A(p)+\Lambda,
\label{2-matrix S-D 2}
\end{eqnarray}
where
\begin{eqnarray}
x(p) & = & 8 + 4p + 8p^2 + p^3,
\nonumber \\
a^{(B^k)}(p) & = & (4+8p+p^2) [B^k] + (8+p) [A B^k] + [A^2 B^k].
\end{eqnarray}
Note that $[{\bf 1}]= \Lambda$
and we used the $Z_2$ symmetry.
One can easily find that eqs.(\ref{2-matrix S-D 1}) for $k = 0, 1, 2$ and 
eq.(\ref{2-matrix S-D 2}) make a closed set
of equations \cite{GN,Staudacher}.
We can eliminate $W^{(B)}_A(p)$, $W^{(B^2)}_A(p)$ and $W^{(B^3)}_A(p)$, 
and have the following fourth order equation of $W_A(p)$:
\begin{equation}
V_A(p)^4+\alpha_3(p) V_A(p)^3+\alpha_2(p) V_A(p)^2+\alpha_1(p) V_A(p)
      +\alpha_0(p) = 0,
\label{45disk1}
\end{equation}
where
\begin{eqnarray}
V_A(p) & = & W_A(p) - x(p),
\nonumber \\
\alpha_3(p) & = & x(p) - 8,
\nonumber \\
\alpha_2(p) & = & 4 - 8x(p) + a^{(\bf{1})}(p),
\\
\alpha_1(p) & = & p - 8 + 4x(p) - 8a^{(\bf{1})}(p) - a^{(B)}(p),
\nonumber \\
\alpha_0(p) & = & - \Lambda + (p-8)x(p) + 4a^{(\bf{1})}(p) + 8a^{(B)}(p) +
                  a^{(B^2)}(p).
\nonumber 
\end{eqnarray} 
We must provide the amplitudes $[A]$, $[A^2]$, $[A B]$,
$[A^2 B]$ and $[A^2 B^2]$ in order to solve eq.(\ref{45disk1}). 
These can be determined by 
the method of orthogonal polynomial (see appendix).
The continuum limit can be carried out by the renormalizaion
$\Lambda = 70-10 a^2 t$ and $p = 2 a \zeta_A$ with the lattice 
spacing $a$ \cite{GM}. Here $70$ is a critical value of $\Lambda$.
Assuming the scaling form of $V_A(p)$ as
\beq
V_A(p)= c_0+c_1 \zeta_A a +c_2 w_A(\zeta_A,t) a^{5/4} +{\cal O}(a^{6/4}),
\label{V_A}
\eeq
and substituting this form into eq.(\ref{45disk1}), we have
the equation of $w_A(\zeta_A,t)$,
\begin{eqnarray}
w_A(\zeta_A, t)^4 -4 t^{5/4} w_A(\zeta_A, t)^2 
+ 2(t^{5/2}-5 t^2 \zeta_A +20 t\zeta^3_A - 16 \zeta^5_A) = 0
\;,
\label{continuum SD-eq}\end{eqnarray} 
and find $(c_0,c_1,c_2)=(0,-2,\pm 2)$. 
By solving eq.(\ref{continuum SD-eq}), we can obtain the continuum 
universal disk amplitude $w_A(\zeta_A,t)$ as
\begin{eqnarray}
w_A(\zeta_A, t) & = & \left(\zeta_A + \sqrt{\zeta^2_A - t} \right)^{5/4} 
              + \left(\zeta_A - \sqrt{\zeta^2_A - t} \right)^{5/4}.
\label{AnoTanshoku}
\end{eqnarray}


We examine next the amplitude $W_{AB}(p,q)$ which is the
prime interest.
Observing that 
$W_{A B}(p, q) = \sum^{\infty}_{k=0} W^{(B^k)}_A(p) q^{-k-1}$,
from eqs.(\ref{2-matrix S-D 1}), (\ref{2-matrix S-D 2})  
we can find the following equation, 
\begin{eqnarray}
W_{A B}(p, q) = \frac{(4+8p+p^2)W_A(q) + (8+p)W^{(B)}_A(q)
                          + W^{(B^2)}_A(q) - W_A(p)}
                     {x(p) - q - W_A(p)}
\;.
\end{eqnarray}
A careful consideration is needed to extract the universal amplitude 
$w_{A B}(\zeta_A, \zeta_B, t)$ from
 this equation. 
 For example let us consider $W_A^{(B^m)}(p)$ for finite $m$.
 The boundary of the corresponding disk involves a part of finite
 lattice length composed of the matrix $B$. 
 Then the contribution from such a part in
 $W_{AB}(p,q)$ turns out to be non-universal.
  In general, any polynomials of $\zeta_A$ and $\zeta_B$
 multiplied by $W_A^{(B^m)}(p)$ or $W_B^{(A^m)}(q)$ are
 non-universal quantities of $W_{AB}(p,q)$.
Polynomials of $\zeta_A$ and $\zeta_B$ are also non-universal.
 We should drop these quantities appropriately to extract a
 universal part of $W_{AB}(p,q)$.  
Using the expansion of $V_A(p)$ (\ref{V_A}) and a similar expansion
of $V_B(q)$, we can find
\beqy
&&W_{AB}(p,q)-2(4+a\zeta_A+a \zeta_B)\left(W_{A}(p)+W_{B}(q)\right)
\nonumber \\
&&\quad 
-\left(W_{A}^{(B)}(p)+W_{B}^{(A)}(q)\right)
\nonumber \\
&&\quad
= -7 -32(\zeta_A+\zeta_B) a -4(\zeta_A^2+4\zeta_A \zeta_B
+\zeta_B^2)a^2
\nonumber \\
&&\quad
 +4\left( 4t^{5/4}+ w_A(\zeta_A,t) w_B(\zeta_B,t)
\right)a^{5/2} +{\cal O}(a^{11/4})
\;.
\eeqy
In the left hand side, we appropriately subtracted 
some non-universal quantities
in advance. 
Moreover we should drop any terms which are analytic in both 
$\zeta_A$ and $\zeta_B$ from the right hand side.
Therefore 
we can read the continuum universal part of $W_{AB}(p,q)$ as
\beq
w_{AB}(\zeta_A,\zeta_B,t)=w_A(\zeta_A,t)w_B(\zeta_B,t)
\;,
\label{wAB}\eeq
where $w_B(\zeta,t) = w_A(\zeta,t)$ from the $Z_2$ symmetry.
It should be noted
that the terms with order higher than $a^{5/4}$ in $V_A(p)$ (\ref{V_A}) 
do not appear in the right hand side of eq.(\ref{wAB}), so that 
$w_{AB}(\zeta_A, \zeta_B, t)$ can
be expressed only in terms of $w_A(\zeta_A, t)$ and $w_B(\zeta_B, t)$.
We will discuss the implication of this fact in sect.4.

\section{Three-matrix model case}
 In this section, we will investigate the disk amplitudes
\beqy
 &&\Wbar_{AB}(p,q)=
\sum_{n,m=0}^{\infty}
\frac{\Lm}{N}\left\langle tr(A^n B^m)\right\rangle
p^{-n-1}q^{-m-1}
\;, \label{3-1} \\
 &&\Wbar_{AC}(p,r)=
\sum_{n,m=0}^{\infty}
\frac{\Lm}{N}\left\langle tr(A^n C^m)\right\rangle
p^{-n-1}r^{-m-1}\;,
\label{3-matrix want}
\eeqy
and their continuum universal parts
$\wbar_{AB}(\zeta_A, \zeta_B, t)$, 
$\wbar_{AC}(\zeta_A, \zeta_C, t)$
in the three-matrix model.
As potentials which describe the $(4,5)$ model, we take
\begin{eqnarray}
U_1(\phi) & = & \frac{111}{16} \phi - \frac{9}{4} \phi^2
                - \frac{1}{3} \phi^3,
\label{3-matrix potential}\\
U_2(\phi) & = & - \frac{3}{4} \phi^2 - \frac{1}{12} \phi^3. 
\nonumber
\label{Action2}
\end{eqnarray}
These can be found by the orthogonal
polynomial method (see appendix).
In order to obtain the amplitudes (\ref{3-1}), (\ref{3-matrix want}), 
we have to calculate 
$\Wbar_{A}^{(B^m C^k)}(p)=\sum^{\infty}_{n=0} [A^n B^m C^k] p^{-(n+1)}$ and 
$\Wbar_{B}^{(A^n C^k)}(q)=\sum^{\infty}_{m=0} [A^n B^m C^k]
q^{-(m+1)}$,
where $[A^n B^m C^k] = \frac{\Lambda}{N} \langle tr (A^n B^m C^k) \rangle$. 

Let us examine $\Wbar_{A}^{(B^m C^k)}(p)$ first.
 Consider the Schwinger-Dyson equations  
\begin{eqnarray}
0 & = & \sum_a \int [dAdBdC] \frac{\partial}{\partial A^a}
            \left\{ tr \left( A^{n} t^a B^m C^k \right) e^{-S(A,B,C)}
            \right\}, 
\nonumber \\
0 & = & \sum_a \int [dAdBdC] \frac{\partial}{\partial B^a}
            \left\{ tr \left( A^{n} t^a       C^k \right) e^{-S(A,B,C)}
            \right\},
\\
0 & = & \sum_a \int [dAdBdC] \frac{\partial}{\partial C^a} 
            \left\{ tr \left( A^{n} B^{m} t^a     \right) e^{-S(A,B,C)}
            \right\}.
\nonumber
\label{3-matrix S-D eq}
\end{eqnarray}
We may write them in the resolvent
representation,
\begin{eqnarray}
0 & = & \left\{\Wbar_A(p)-y(p) \right\} \Wbar^{(B^m C^k)}_A(p) 
      + \Wbar^{(B^{m+1} C^k)}_A(p) + a^{(B^m C^k)}(p), 
\label{3mx SD A 1}\\
0 & = & - \frac{3}{2} \Wbar^{(B C^k)}_A(p) - \frac{1}{4} \Wbar^{(B^2 C^k)}_A(p)
        - \Wbar^{(C^{k+1})}_A(p) - p \Wbar^{(C^k)}_A(p) + [A^k],
\label{3mx SD A 2}\\
0 & = & \frac{111}{16}\Wbar^{(B^m)}_A(p) - \frac{9}{2} \Wbar^{(B^m C)}_A(p)
        - \Wbar^{(B^m C^2)}_A(p) - \Wbar^{(B^{m+1})}_A(p),
\label{3mx SD A 3}
\end{eqnarray}
where
\begin{eqnarray}
y(p) & = & \frac{111}{16} - \frac{9}{2} p - p^2,
\nonumber \\
a^{(B^m C^k)}(p) & = & -(\frac{9}{2} + p) [B^m C^k] - [A B^m C^k]. 
\nonumber 
\nonumber
\end{eqnarray}
Here $[{\bf 1}] = \Lambda$ and we used the $Z_2$ symmetry.
One can find that eqs.(\ref{3mx SD A 1}) for $(m,k)=(0,0),(0,1),(1,0),(1,1)$, 
eqs.(\ref{3mx SD A 2}) for $k=0,1$ and eq.(\ref{3mx SD A 3}) 
 for $m=0$ are independent and make a closed set of equations\cite{AIT}. 
By eliminating
$\Wbar^{(B)}(p)$, $\Wbar^{(C)}(p)$, $\Wbar^{(B^2)}(p)$, $\Wbar^{(B C)}(p)$, 
$\Wbar^{(C^2)}(p)$ and $\Wbar^{(B^2 C)}(p)$,
we obtain the following fifth order 
equation of
$\Wbar_A(p)$,
\begin{equation}
U_A(p)^5 + \overline{\alpha}_4(p) U_A(p)^4
 + \overline{\alpha}_3(p) U_A(p)^3 + \overline{\alpha}_2(p) U_A(p)^2 
    + \overline{\alpha}_1(p) U_A(p) + \overline{\alpha}_0(p) = 0.
\label{3mx SD A}\end{equation}
Here
\begin{eqnarray}
U_A(p) & = & \Wbar_A(p) - y(p),
\nonumber \\
\overline{\alpha}_4(p) & = & -12 + y(p), 
\nonumber \\
\overline{\alpha}_3(p) & = & 18 + 8p + a^{(1)}(p) -12y(p),
\nonumber \\
\overline{\alpha}_2(p) & = & 92 - 48p - 12a^{(1)}(p) - a^{(B)}(p)
                  - 4\Lambda + (18+8p)y(p),
\nonumber \\
\overline{\alpha}_1(p) & = & -111 - 72p + 16p^2 + (18+4p)a^{(1)}(p)
                  + 6a^{(B)}(p) - 4a^{(C)}(p) 
\nonumber \\
                && + 24\Lambda + (92-48p)y(p),
\label{3mx SD 5}\\
\overline{\alpha}_0(p) & = & (92-24p)a^{(1)}(p) + (18-4p)a^{(B)}(p)
                  + 24a^{(C)}(p) + 4a^{(B C)}(p)
\nonumber \\
                && + (72-16p)\Lambda + 16[A] + (-111-72p+16p^2)y(p).
\nonumber
\end{eqnarray}
The amplitudes
$[A]$, $[B]$, $[A B]$,
$[A C]$ and $[A B C]$
are determined by the orthogonal polynomial method 
(see appendix).
With the renormalization, 
$\Lambda = 35 -\frac 52 a^2 t$ and $p = \frac 32 a \zeta_A$,
 we assume the scaling behavior of $U_A(p)$  as
\begin{eqnarray}
U_A(p) = \bar{c}_0 + \bar{c}_1 \zeta_A a + \bar{c}_2 \wbar_A(\zeta_A, t) a^{5/4} + {\cal O} (a^{6/4}).
\label{U_A}
\label{3mx scaling A}
\end{eqnarray}
Substituting eq.(\ref{3mx scaling A}) into eq.(\ref{3mx SD A}) 
and after similar calculation in sect.2,
we find that $(\bar{c}_0,\bar{c}_1,\bar{c}_2)=(0,2,\pm 2/3)$ and
\begin{eqnarray}
\wbar_A(\zeta_A, t) = \left(\zeta_A + \sqrt{\zeta_A^2 - t} \right)^{5/4} + 
            \left(\zeta_A - \sqrt{\zeta_A^2 - t} \right)^{5/4}.
\end{eqnarray}
As expected, this coincides with the result for 
$w_A(\zeta_A, t)$ in the two-matrix model, eq.(\ref{AnoTanshoku}).

Next let us examine $\Wbar_B(q)$. In this case, we found that 
ten Schwinger-Dyson equations are needed. For example 
let us consider the following ten equations:
\begin{eqnarray}
0 & = & \sum_a \int [dAdBdC] \frac{\partial}{\partial B^a} 
        \left\{
              tr \left(  t^a B^{n} \right) e^{-S(A,B,C)} 
        \right\},
\nonumber \\
0 & = & \sum_a \int [dAdBdC] \frac{\partial}{\partial B^a} 
        \left\{
              tr \left(A t^a B^{n} \right) e^{-S(A,B,C)} 
        \right\},
\nonumber \\
0 & = & \sum_a \int [dAdBdC] \frac{\partial}{\partial B^a} 
        \left\{
              tr \left(C t^a A B^{n} \right) e^{-S(A,B,C)} 
        \right\},
\nonumber \\
0 & = & \sum_a \int [dAdBdC] \frac{\partial}{\partial B^a} 
        \left\{
              tr \left(C A t^a B^{n} \right) e^{-S(A,B,C)} 
        \right\},
\nonumber \\
0 & = & \sum_a \int [dAdBdC] \frac{\partial}{\partial B^a} 
        \left\{
              tr \left(A B C t^a B^{n} \right) e^{-S(A,B,C)} 
        \right\},
\\
0 & = & \sum_a \int [dAdBdC] \frac{\partial}{\partial A^a} 
        \left\{
              tr \left( t^a B^{n} \right) e^{-S(A,B,C)} 
        \right\},
\nonumber \\
0 & = & \sum_a \int [dAdBdC] \frac{\partial}{\partial A^a} 
        \left\{
        tr \left( \left(C A t^a - C t^a A \right) B^{n} \right) e^{-S(A,B,C)} 
        \right\},
\nonumber \\
0 & = & \sum_a \int [dAdBdC] \frac{\partial}{\partial A^a} 
        \left\{
              tr \left(C B t^a B^{n} \right) e^{-S(A,B,C)} 
        \right\},
\nonumber \\
0 & = & \sum_a \int [dAdBdC] \frac{\partial}{\partial A^a} 
        \left\{
        tr \left( \left(A C A t^a - A C t^a A \right) B^{n} \right)
               e^{-S(A,B,C)} 
        \right\},
\nonumber \\
0 & = & \sum_a \int [dAdBdC] \frac{\partial}{\partial A^a} 
        \left\{
        tr \left( \left(C B A t^a - C B t^a A \right) B^{n} \right)
               e^{-S(A,B,C)} 
        \right\}.
\nonumber
\end{eqnarray}
 In the resolvent representation, we have
\begin{eqnarray}
0 & = & \left\{ \Wbar_B(q)-z(q) \right\}
 \Wbar_B(q) + 2\Wbar^{(A)}_B(q) + b^{({\bf 1})}(q),
\nonumber \\
0 & = & \left\{ \Wbar_B(q)-z(q) \right\} \Wbar^{(A)}_B(q) 
        + \Wbar^{(A^2)}_B(q) + \Wbar^{(A C)}_B(q) + b^{(A)}(q), 
\nonumber \\
0 & = & \left\{ \Wbar^{(A)}_B(q) \right\}^2
        + \frac{3}{2} \Wbar^{(A B C)}_B(q) +\frac{1}{4} \Wbar^{(A B^2 C)}_B(q)
        + 2\Wbar^{(A^2 C)}_B(q), 
\nonumber
\label{3mx SD B 2}\\
0 & = & \left\{ \Wbar_B(q)-z(q) \right\} \Wbar^{(A C)}_B(q) 
        + \Wbar^{(A^2 C)}_B(q) + \Wbar^{(A C A)}_B(q) + b^{(A C)}(q), 
\nonumber\\
0 & = & \left\{ \Wbar_B(q)-z(q) \right\} \Wbar^{(A B C)}_B(q) 
        + \Wbar^{(A B C A)}_B(q) + \Wbar^{(A^2 B C)}_B(q)
\nonumber \\
    && \qquad + [A] \Wbar^{(A)}_B(q) + b^{(A B C)}(q), 
\label{3mx SD B 1} \\
0 & = & \left(\frac{111}{16} - q \right)\Wbar_B(q)
   - \frac{9}{2} \Wbar^{(A)}_B(q) 
        - \Wbar^{(A^2)}_B(q) + \Lambda,
\nonumber\\
0 & = & q \Wbar^{(A C)}_B(q) - \Wbar^{(A B C)}_B(q) - [A C],
\nonumber\\
0 & = & \left( \frac{111}{16}q - q^2 \right) \Wbar^{(A)}_B(q) 
        - \frac{9}{2}\Wbar^{(A B C)}_B(q) - \Wbar^{(A^2 B C)}_B(q)
        - \left( \frac{111}{16} - q \right) [A] + [A B],
\nonumber\\
0 & = & [A C] \Wbar_B(q) - [A] \Wbar^{(A)}_B(q) - q\Wbar^{(A C A)}_B(q)
        + \Wbar^{(A B C A)}_B(q) + [A^2 C],
\nonumber\\
0 & = & q\Wbar^{(A B C)}_B(q) - \Wbar^{(A B^2 C)}_B(q) - [A B C]
\;,
\nonumber
\end{eqnarray}
where
\begin{eqnarray}
z(q) & = & -\frac{3}{2}q-\frac{1}{4}q^2,
\nonumber\\
b^{(A^n B^m C^k)}(q) & = & -(\frac{3}{2} + \frac{1}{4} q) [A^n B^m C^k]
                        - \frac{1}{4} [B A^n B^m C^k]
\;.
\end{eqnarray}
We can find that these are independent and make a closed set of equations.
The fourth order equation which determines $\Wbar_B(q)$ is obtained as:
\begin{eqnarray}
U_B(q)^4 + \overline{\beta}_2(q) U_B(q)^2 +\overline{\beta}_0(q) = 0,
\label{3mx U_B}
\end{eqnarray}
where
\begin{eqnarray}
U_B(q) &=& \Wbar_B(q) -z(q)-\frac{9}{2}.
\nonumber
\end{eqnarray}
The coefficients of this equation are given by
\begin{eqnarray}
\overline{\beta}_2(q) & = &-\frac{273}{4} -3\Lambda - \frac{[B]}{2}
                + \left(\frac{35}{2} -\frac{\Lambda}{2} \right)q 
                - \frac{3}{4} q^3 
                - \frac{1}{16} q^4,
\nonumber\\
\overline{\beta}_0(q) & = &   972 + \frac{135}{8}
     \Lambda + \frac{9}{4} \Lambda^2
                        - 46[A] - \frac{51}{16}[B] + \frac{3}{4}\Lambda [B]
                        - 6[A C] + \frac{[B]^2}{16}
                        - 9[A B] - [A B C]
\nonumber\\
                &&+ \left\{-729 + \frac{81}{16} \Lambda 
                               + \frac{3}{4} \Lambda^2
                               - [A C]
                               + \frac{19}{8} [B]
                               + \frac{1}{8} \Lambda [B]
                               + \frac{3}{2} [A B] 
                    \right\}q
\nonumber\\
                &&+ \left\{54 + \frac{19}{4} \Lambda + \frac{\Lambda^2}{16}
                           + 3[A] + \frac{9}{16}[B] + \frac{[A B]}{4}
                    \right\}q^2
\nonumber\\
                &&+ \left\{36 + \frac{9}{16} \Lambda + \frac{[A]}{4}
                    \right\}q^3
                  - \frac{q^5}{4}.
\end{eqnarray}
With the renormalization $q=2 a \zeta_B$,
solving eq.(\ref{3mx U_B}) directly, we can find the 
the disk amplitude $U_B(q)$ as
\beq
U_B(q)=\pm~\wbar_B(\zeta_B,t)a^{5/4} 
+{\cal O}(a^{6/4})
\;,
\label{3.17}
\eeq
where
$\wbar_B(\zeta_B, t)$ coincides with 
$\wbar_A(\zeta_B, t)$.


 Now let us turn to the calculation of $\Wbar_{AB}(p, q)$, 
$\Wbar_{AC}(p, r)$. From
eq.(\ref{3mx SD A 1}) for $k=0$, one can obtain the relation
\begin{eqnarray}
\Wbar_{A B}(p, q) = \frac{(\frac{9}{2}+p)\Wbar_B(q)
 + \Wbar^{(A)}_B(q) + \Wbar_A(p)}
                        { \Wbar_A(p)-y(p)+q }.
\label{SD AB}
\end{eqnarray}
By combining eq.(\ref{SD AB}) and the first equation of (\ref{3mx SD B
1}), $\Wbar_{AB}(p,q)$
can be expressed in terms of $\Wbar_A(p)$ and $\Wbar_B(q)$.
In order to extract a universal part, we must drop polynomials of
$\zeta_A$ and $\zeta_B$ multiplied by $\Wbar_A^{(B^m C^k)}(p)$ or 
$\Wbar_B^{(A^n C^k)}(q)$ 
as well as polynomials of both $\zeta_A$ and $\zeta_B$
appropriately, because of the same reason
as stated in sect.2.
Using the expressions (\ref{U_A}) and (\ref{3.17}),
we can find
\beqy
&&\Wbar_{AB}(p,q)-\frac{9}{16}\Wbar_A(p) 
 -\frac{3}{4}\Wbar_{B}(q) 
\nonumber \\
&&\quad=
-\frac{1607}{256} +\frac{9}{64}(11\zeta_A +8 \zeta_B) a
\label{WbarAB}\\
&&\quad-\frac{1}{8} \frac{\wbar_A(\zeta_A,t)^2 +2 \wbar_A(\zeta_A,t)
\wbar_B(\zeta_B,t) +2 \wbar_B(\zeta_B,t)^2 -2 t^{5/4} }
{\zeta_A +\zeta_B} a^{3/2}
+{\cal O}(a^{7/4})
\; . \nonumber
\eeqy
In the left hand side, we subtracted some non-universal quantities
in advance appropriately.
Moreover we should drop first and second terms in the right hand side,
because they are polynomials of both $\zeta_A$ and $\zeta_B$.
From this equation, we can find 
the continuum universal
disk amplitude $\wbar_{AB}(\zeta_A,\zeta_B,t)$ as
\beq
\wbar_{AB}(\zeta_A,\zeta_B,t)=
\frac{\wbar_A(\zeta_A,t)^2
+2\wbar_A(\zeta_A,t)\wbar_B(\zeta_B,t)+2\wbar_B(\zeta_B,t)^2
-2t^{5/4} }{\zeta_A+\zeta_B}
\;.
\label{wbarAB}
\eeq
We can observe that terms with order higher than $a^{5/4}$ in $\Wbar_A(p)$
and $\Wbar_B(q)$ do not appear in the right hand side of eq.(\ref{wbarAB}).
Thus $\wbar_{AB}(\zeta_A, \zeta_B, t)$ is expressed only in terms of 
$\wbar_A(\zeta_A, t)$ and $\wbar_B(\zeta_B, t)$.

Next let us consider $\Wbar_{AC}(p,r)$.
From eqs.(\ref{3mx SD A 1}) for $m=0, 1$
and eq.(\ref{3mx SD A 2}), we can obtain the equations
\begin{eqnarray}
\Wbar_{A C}(p,r) = \frac{(\frac{9}{2}+p)\Wbar_C(r) + \Wbar^{(A)}_C(r) 
                                 - \Wbar^{(B)}_{A C}(p, r)}
                        {\Wbar_A(p)-y(p)}
\;,
\end{eqnarray}
\begin{eqnarray}
\Wbar^{(B)}_{A C}(p, 0, r) = 
   \frac{(\frac{9}{2}+p)\Wbar^{(B)}_C(r) + \Wbar^{(AB)}_C(r) 
                                 - \Wbar^{(B^2)}_{A C}(p, r)}
                        {\Wbar_A(p)-y(p)}
\;,
\end{eqnarray}
\begin{eqnarray}
\frac{3}{2} \Wbar^{(B)}_{A C}(p, r) + \frac{1}{4} \Wbar^{(B^2)}_{A C}(p, r)
  +(p+r)\Wbar_{A C}(p, r) - \Wbar_A(p) - \Wbar_C(r) = 0
\;,
\end{eqnarray}
respectively. By combining these, we can express $\Wbar_{AC}(p, r)$
in terms of $\Wbar_A(p)$ and $\Wbar_C(r)$.
Note that $\Wbar_C(r)= \Wbar_A(r)$ because of the $Z_2$ symmetry.   
Using the expression (\ref{U_A}) and (\ref{3.17}), we find
\beqy
&&\Wbar_{AC}(p,r)-\frac{41}{5}\left(\Wbar_A^{(C^2)}(p) +
\Wbar_C^{(A^2)}(r)\right)
-\frac{801}{112}\left(\Wbar_A^{(C)}(p)+\Wbar_C^{(C)}(r)\right)
\nonumber \\
&&\quad
-\left(\frac{90}{7}- \frac{11}{32}a\zeta_C \right)\Wbar_A(p) 
-\left(\frac{90}{7}- \frac{11}{32}a\zeta_A \right)\Wbar_C(r)
\nonumber \\
&&\quad=
-\frac{44649}{7168} +\frac{99}{32}(\zeta_A+\zeta_C) a
-\frac{1}{3584} \left(-23200 t+1377\zeta_A^2 +8120\zeta_A\zeta_C
+1377\zeta_C^2 \right) a^2
\nonumber \\
&&\quad
+\left(-\frac{73}{28} t^{5/4} +\frac{1}{9}\wbar_A(\zeta_A,t)
\wbar_C(\zeta_C,t)\right) a^{5/2} +{\cal O}(a^{11/4})
\;.
\eeqy
Here we subtracted some non-universal quantities in advance 
appropriately from
$\Wbar_{AC}(p, r)$.
The first, second, third and $t^{5/4}$ terms in the right hand side
should be dropped, because they are polynomials of both
$\zeta_A$ and $\zeta_B$.
We can read, therefore, the continuum universal
disk amplitude 
$\wbar_{AC}(\zeta_A,\zeta_C,t)$ as
\beq
\wbar_{AC}(\zeta_A,\zeta_C,t)=\wbar_A(\zeta_A,t)
\wbar_C(\zeta_C,t)
\;.
\label{wbarAC}\eeq


\section{Comparison and interpretation}
 In the previous two sections, we obtained the disk amplitudes
 with heterogeneous boundaries 
 $w_{AB}(\zeta_A, \zeta_B, t)$, $\wbar_{AB}(\zeta_A, \zeta_B, t)$
 and $\wbar_{AC}(\zeta_A, \zeta_C, t)$ 
 ( eqs.(\ref{wAB}), (\ref{wbarAB}) and (\ref{wbarAC}) respectively).
 In this section, we will compare them and
 provide 
 a geometrical interpretation of these amplitudes. 
 In this and the next sections, we denote the matrices $A$, $B$ and $C$
 in the three matrix
 model as $\Abar$, $\Bbar$ and $\Cbar$ respectively,  in order to
 distinguish from those in the two-matrix model. We will refer to
 a part of boundary which is composed of the matrix $A$
 as ``boundary $A$''  and so on.

 From eqs.(\ref{wAB}) and (\ref{wbarAC}), we observe that $w_{AB}$ and 
$\wbar_{\Abar \Cbar}$
 have exactly the same form.
  We can consider that boundaries $A$ and $B$ correspond to
 boundaries $\Abar$ and $\Cbar$ respectively.
In the case of loops with homogeneous matter states,
this correspondence
 is natural from the view point of the orthogonal polynomial
 method. In the case of heterogeneous boundaries, however,
 the method of the orthogonal polynomial cannot be applied
and this correspondence is not so trivial.

  From eqs.(\ref{wbarAB}) and (\ref{wbarAC}), 
 we can observe that $\wbar_{\Abar \Bbar}$ 
 and $\wbar_{\Abar \Cbar}$
 have quite different forms.
But $\wbar_{\Abar \Bbar}$ obtained by eq.(\ref{wbarAB})
 has a very similar form
 to the disk amplitude $w^{(I)}_{AB}$ in the case of the Ising model 
realized by the two-matrix model \cite{IIKMNS,SY}:
\beq
w^{(I)}_{AB}(\zeta_A,\zeta_B,t)=
\frac{w^{(I)}(\zeta_A,t)^2
+w^{(I)}(\zeta_A,t)w^{(I)}(\zeta_B,t)+w^{(I)}(\zeta_B,t)^2
-3t^{4/3} }{\zeta_A+\zeta_B}
\;.
\label{wIAB}
\eeq
Here $w^{(I)}$ is given by
\beq
w^{(I)}(\zeta,t)=\left( \zeta +\sqrt{\zeta^2-t} \right)^{4/3}
+\left( \zeta -\sqrt{\zeta^2-t} \right)^{4/3}
\;.
\eeq
Note that, however,
there is no symmetry under interchange of $\Abar$ and $\Bbar$
in eq. (\ref{wbarAB}).

 For the sake of discussing why $\wbar_{\Abar \Bbar}$ 
and $\wbar_{\Abar \Cbar}$
 have so different forms,
it is useful to
 consider the inverse Laplace transformed amplitudes.
 Let us denote the inverse Laplace transformed amplitudes of
 $w_A(\zeta_A,t)$ as $\cW_A(\ell_A)$ etc.
 For example, $w_{AB}(\zeta_A,\zeta_B,t)$ and
 $\cW_{AB}(\ell_A,\ell_B)$ are related by the equation
\beq
w_{AB}(\zeta_A,\zeta_B,t)
={\cal L}_A {\cal L}_B \left[\cW_{AB}(\ell_A,\ell_B)\right]
=\int_0^{\infty}d \ell_A \int_0^{\infty}d \ell_B 
e^{-\ell_A \zeta_A-\ell_B \zeta_B} \cW_{AB}(\ell_A,\ell_B).
\eeq
 Here $\cW_{AB}(\ell_A,\ell_B)$ represents a disk amplitude 
 where length of each part of the boundary is fixed.
First we easily obtain the relations:
\begin{eqnarray}
\cW_{AB}(\ell_A,\ell_B)
&=&\cW_A(\ell_A) \cW_B(\ell_B)
\label{4-4}\;,
\\
\cWbar_{\Abar \Cbar}(\ell_{\Abar},\ell_{\Cbar})
&=&\cWbar_{\Abar}(\ell_{\Abar})  \cWbar_{\Cbar} (\ell_{\Cbar})
\label{4-5} \;.
\end{eqnarray}

 As for $\cWbar_{\Abar \Bbar}$, we use the following formulas of the inverse
 Laplace transformation;
\beq
{\cal L}^{-1}\left[ \frac{1}{\zeta +a} \right]
=e^{-a\ell}
\;,
\label{formula1}\eeq
\beq
{\cal L}^{-1}\left[ e^{-\zeta a}F(\zeta) \right]
=\theta(\ell-a)~f(\ell-a)
\;,
\label{formula2}\eeq
\beq
{\cal L}^{-1}\left[ \frac{F(\zeta)}{\zeta +a} \right]
=e^{-a\zeta} \int_0^{\ell} dx e^{ax} f(x)
\;,
\label{formula3}\eeq
 where $F(\zeta)$ denotes the image of the Laplace transformation of
 $f(\ell)$.
Using the formulas (\ref{formula1}) and (\ref{formula2}),
we obtain the relation:
\beqy
&&{\cal L}_{\Abar}^{-1}{\cal L}_{\Bbar}^{-1}\left[
 \frac{\wbar_{\Abar}(\zeta_{\Abar},t)^2}{\zeta_{\Abar} +\zeta_{\Bbar}} \right]
={\cal L}_{\Abar}^{-1} \left[ 
 e^{-\zeta_{\Abar}\ell_{\Bbar}} \wbar_{\Abar}(\zeta_{\Abar},t)^2 \right]
\nonumber \\
&&\qquad
=\theta(\ell_{\Abar}-\ell_{\Bbar})~{\cal L}_{\Abar}^{-1} \left[
\wbar_{\Abar}(\zeta_{\Abar},t)^2 \right]
\nonumber \\
&&\qquad
=\theta(\ell_{\Abar}-\ell_{\Bbar}) \int_0^{\ell_{\Abar}-\ell_{\Bbar}}
d\ell \cWbar_{\Abar}(\ell) \cWbar_{\Abar}(\ell_{\Abar}-\ell_{\Bbar}-\ell)
\;.
\label{cWbarAB1}\eeqy
We also have
\beqy
&&{\cal L}_{\Abar}^{-1}{\cal L}_{\Bbar}^{-1}\left[
 \frac{\wbar_{\Abar}(\zeta_{\Abar},t)\wbar_{\Bbar}(\zeta_{\Bbar},t)}
  {\zeta_{\Abar} +\zeta_{\Bbar}} \right]
={\cal L}_{\Bbar}^{-1} \left[ 
 \int_0^{\ell_{\Abar}} d\ell e^{-\zeta_{\Bbar}(\ell_{\Abar}-\ell)}
 \cWbar_{\Abar}(\ell) \wbar_{\Bbar}(\zeta_{\Bbar},t) \right]
\nonumber \\
&&\qquad
=\int_0^{\ell_{\Abar}} d\ell \cWbar_{\Abar}(\ell)
\theta(\ell_{\Bbar}-\ell_{\Abar}+\ell)\cWbar_{\Bbar}
(\ell_{\Bbar}-\ell_{\Abar}+\ell)
\;,
\label{cWbarAB2}\eeqy
by using the formulas (\ref{formula3}) and (\ref{formula2}).
From (\ref{formula1}) we have
\beq
{\cal L}_{\Abar}^{-1}{\cal L}_{\Bbar}^{-1}\left[
 \frac{1}{\zeta_{\Abar}+\zeta_{\Bbar}} \right]
={\cal L}_{\Bbar}^{-1} \left[ e^{-\zeta_{\Bbar}\ell_{\Abar}} \right]
=\delta(\ell_{\Abar}-\ell_{\Bbar})
\;.
\label{cWbarAB3}\eeq
Collecting eqs.(\ref{cWbarAB1}) - (\ref{cWbarAB3}) together,
we obtain the expression for $\cWbar_{\Abar \Bbar}$, 
\beqy
&&\cWbar_{\Abar \Bbar}(\ell_{\Abar} \ell_{\Bbar})
=
\theta(\ell_{\Abar}-\ell_{\Bbar})\int_0^{\ell_{\Abar}-\ell_{\Bbar}} d\ell~
\cWbar_{\Abar}(\ell)\cWbar_{\Abar}(\ell_{\Abar}-\ell_{\Bbar}-\ell)
\nonumber \\
&&\qquad+
2\int_0^{{\rm min}(\ell_{\Abar},\ell_{\Bbar})}d\ell~ 
\cWbar_{\Abar}(\ell_{\Abar}-\ell)\cWbar_{\Bbar}(\ell_{\Bbar}-\ell)
\nonumber \\
&&\qquad+
2\;\theta(\ell_{\Bbar}-\ell_{\Abar})\int_0^{\ell_{\Bbar}-\ell_{\Abar}} d\ell~
\cWbar_{\Bbar}(\ell)\cWbar_{\Bbar}(\ell_{\Bbar}-\ell_{\Abar}-\ell)
\nonumber \\
&&\qquad
-2 t^{5/4} \delta(\ell_{\Abar}-\ell_{\Bbar}) \;.
\label{cWbarAB}\eeqy

\begin{figure}
\begin{center}
  \epsfbox{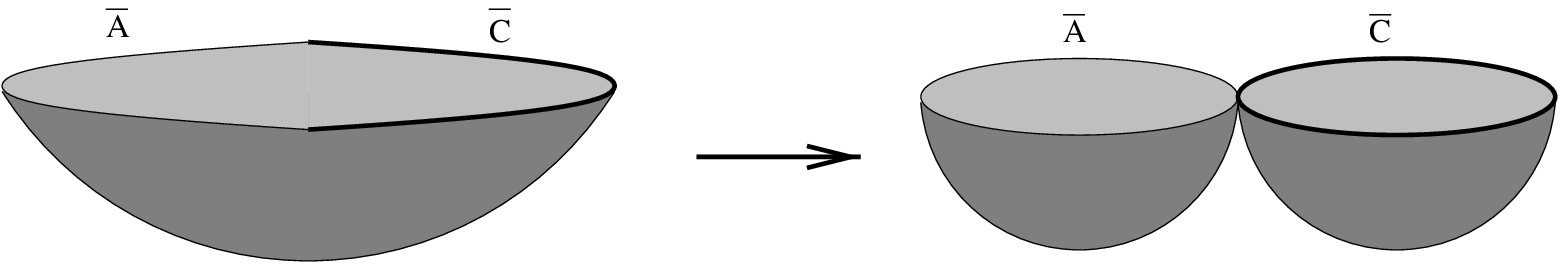}
  \caption{The original loop composed of two different parts of boundary
           splits into two loops each of which has
           homogeneous matter configurations.}
  \label{fig1}
\end{center}
\begin{center}
  \epsfbox{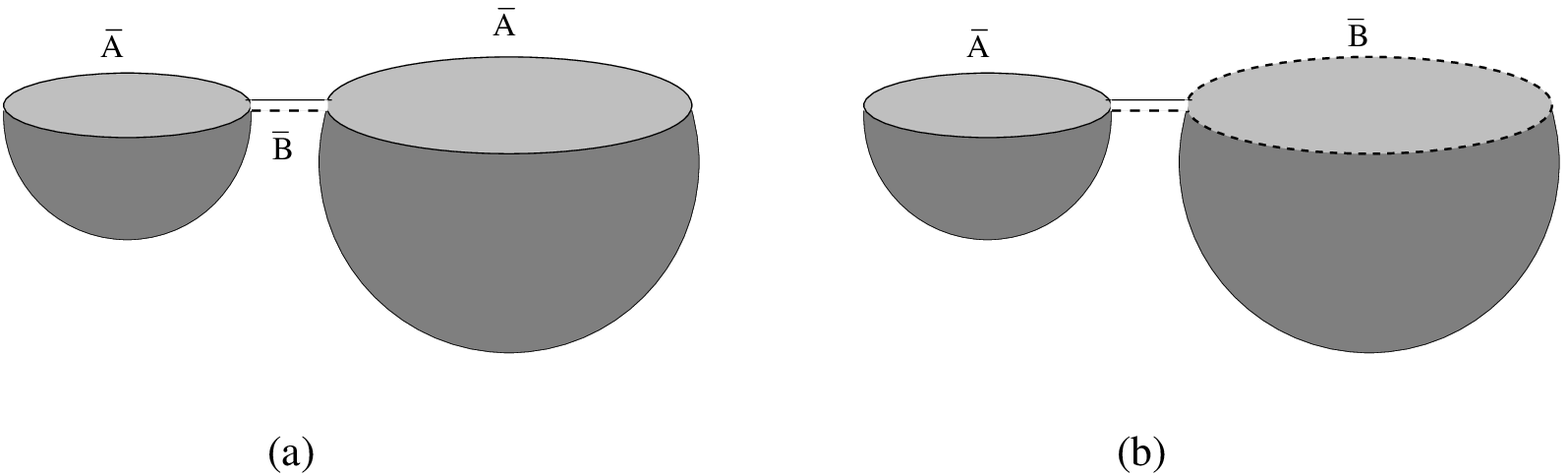}
  \caption{Due to the sticking of two different kinds of boundaries,
           the original loop  splits into two loops with
           homogeneous matter configurations.}
  \label{fig2}
\end{center}
\end{figure}

Now let us consider the geometrical  meaning of
eqs.(\ref{4-5}) and (\ref{cWbarAB}).
As for eq.(\ref{4-5}), it is easy to understand that
a loop composed of boundary $\bar{A}$ and $\bar{C}$
splits into two loops with
different homogeneous matter states (see fig.\ref{fig1}).
On the other hand,
the first term in the right hand side of eq.(\ref{cWbarAB})
represents the configuration depicted in fig.\ref{fig2}(a).
All region of the boundary $\bar{B}$ is stuck to 
the boundary $\bar{A}$, and the
original loop also splits into two loops with homogeneous
matter states. 
Likewise the second term in eq.(\ref{cWbarAB}) 
corresponds to the case
in fig.\ref{fig2}(b). 
Parts of boundaries $\bar{A}$ and $\bar{B}$ are stuck each other,
so that the original loop splits into two loops with homogeneous
matter states.
The fourth term represents the contribution from the case where 
the boundaries $\bar{A}$ and $\bar{B}$ are stuck completely.
From this geometrical picture, 
we can conclude that the original loop, in fact, splits into
two loops with homogeneous matter states.

Next we consider the reason why 
there is a difference between $\cWbar_{\bar{A} \bar{B}}$ and
$\cWbar_{\bar{A} \bar{C}}$ from this geometrical point of view.
In the case of $\cW_{\bar{A}\bar{B}}$ the boundaries $\bar{A}$
and $\bar{B}$ stick each other.
On the other hand,
the boundaries $\bar{A}$ and $\bar{C}$ are connected
at only one point
in $\cW_{\bar{A}\bar{C}}$.
This difference can be accounted for as follows.
The $(4,5)$ minimal conformal model has three degrees of matter freedom 
labeled by the points of the $A_3$ Dynkin diagram.
We can interpret that the matrices $\bar{A}$ and  $\bar{C}$ correspond
to  the ends of the diagram and $\bar{B}$ to
the middle point.
The boundaries $\bar{A}$ and $\bar{B}$
stick each other,
because 
the corresponding states of $\bar{A}$ and $\bar{B}$ interact
directly as opposed to $\bar{A}$ and $\bar{C}$.
In the case of the two-matrix model eq.(\ref{4-4}), 
the matrices $A$ and $B$ correspond to the ends of
the Dynkin diagram. The boundaries $A$ and $B$, therefore,
do not stick each other.

\section{Summary}
In this paper, we have considered the $(4,5)$ minimal model coupled to
2d gravity described by both the two- and the three-matrix models.
We have calculated the disk amplitudes with non-trivial boundary conditions
for the matter configurations, 
and have shown explicitly
the relation among the matrices of these two models.
A geometrical interpretation of the resulting 
amplitudes have also been obtained.
In the process of our calculation in the three-matrix model,
we found that seven and ten Schwinger-Dyson equations make
two closed sets. These two sets of equations determine
the disk amplitudes $\Wbar_{\bar{A}}(p)$ and 
$\Wbar_{\bar{B}}(q)$ respectively.

We obtained the universal disk amplitudes 
$w_{AB}$, $\wbar_{\Abar \Bbar}$ and $\wbar_{\Abar \Cbar}$,
whose boundaries composed of two arcs of finite length 
with different matter states, 
as 
eqs.(\ref{wAB}), (\ref{wbarAB}) and (\ref{wbarAC}) respectively.  
We learned that
the matrices $A$ and $B$ in the two-matrix model
correspond to the matrices $\Abar$ and $\Cbar$ 
in the three-matrix model. 
 
The geometrical meaning of these results is that
the loop of the disk $\cWbar_{\Abar \Bbar}$ or $\cWbar_{\Abar \Cbar}$
splits into two loops
with homogeneous matter states.
Only these configurations contribute to the amplitudes.

In this paper, we only studied disk amplitudes with two arcs. 
It is straightforward to generalize our calculation to disk
amplitudes
with more than two arcs. 
Using the Schwinger-Dyson equations, we can compute recursively the disk
amplitudes with $n$ arcs from those with 
smaller numbers of $n$ arcs.
 
What extent can we apply the technique in this paper 
to other matrix models?
The point is whether we can construct closed sets of equations which
determine the disk amplitudes with homogeneous boundaries
(ex. $\Wbar_{\bar{A}}(p)$ and $\Wbar_{\bar{B}}(q)$).
In the three-matrix model of closed chain type, we can
find that this is the case\footnote{For the action $S=N/\Lambda
tr\{V(A)+V(B)+V(C)-AB-BC-CA\}$ with third order potential $V$,
we can find that eight Schwinger-Dyson equations make a closed set.
 These leads to a fourth order equation.}.
We believe that the Schwinger-Dyson technique could be applied 
successfully to the multi-matrix models of closed chain type
as well as open ones.
 
We believe that the geometrical picture in this paper can be extended to
the cases of 2d surface with many loops and handles.
This expectation is natural because the interaction of boundaries 
is a local phenomenon and does not depend on the global nature of surfaces.
This picture must be applied also to  
the case of general unitary minimal matter. 
Let us consider the $(n+1,n+2)$ unitary model coupled to 2d gravity
realized by the n-matrix chain model.
The matrices naturally correspond to the matter degrees of freedom
labelled by the points of $A_{n}$ Dynkin diagram.
 We expect that two parts of the loops stick each other 
if the two corresponding matter states are adjacent in the Dynkin
diagram.  Due to this mechanism, heterogeneous loops  
must reduce to homogeneous ones. 
This phenomenon may be related to the formulation of
the non-critical string field theory\cite{IK, IIKMNS}, which is constructed in 
the limited space of loops with simple matter configurations.


\section*{Acknowledgment}

We would like to express our gratitude to Prof. M. Ninomiya,
Prof. Y. Matsuo, Prof. H. Kunitomo
and Prof. M. Fukuma for warmful encouragements.
We are grateful to Prof. H. Kunitomo, Prof. M. Fukuma and Dr. F. Sugino
for useful discussions.
Thanks are also due to Prof. H. Kunitomo and Prof. K. Itoh for
careful readings of the manuscript.
This work is supported in part by the Grant-in-Aid for Scientific
Research (2690, 5108) from the Ministry of Education, Science and Culture.

\appendix
\section{Orthogonal Polynomial Method}

In this appendix, we derive the critical potentials 
(\ref{2-matrix potential}), (\ref{3-matrix potential})
and critical values of $\Lambda$ by following ref.\cite{DKK}.
We also show how to evaluate 
$\frac{\Lambda}{N} \langle tr A^n \rangle$ etc.,
which are necessary to solve the Schwinger-Dyson equations in the
text.
We show the details of the calculation by restricting our attention to
the tricritical Ising model; $(p, q) = (4, 5)$.


First, we consider the two-matrix model.
The potential $U(\phi)$ in the action (\ref{2-matrix model Action})
is an arbitrary polynomial;
\begin{eqnarray}
U(\phi) & = & \sum_{k=1}^p \frac{g_k}{k} {\phi}^k.
\end{eqnarray}
The integral over 
matrix elements can be converted into the one over the eigenvalues:
\begin{eqnarray}
Z(\Lambda) & = & \int \prod_{i=1}^N dx_i dy_i \Delta (x) \Delta (y)
e^{- \sum_{i=1}^N S(x_i,y_i)},
\\
S(x_i,y_i) & = & \frac{N}{\Lambda} 
            \left\{U(x_i)+U(y_i)-x_i y_i \right\},
\nonumber\\
\Delta(x) & = & \prod_{i<j} (x_i-x_j)^2.
\nonumber
\end{eqnarray}
By introducing the orthogonal polynomials $\Pi_n (x)$ which satisfy
\begin{eqnarray}
\langle m | n \rangle & \equiv & \int dx dy 
                                 e^{- S(x,y)}
                                 \Pi_m (y) \Pi_n (x)
 =  \delta_{mn},
\end{eqnarray}
we denote the matrix elements as
\begin{eqnarray}
X_{mn} & = & \langle m | x | n \rangle, 
\hspace{1cm}
P_{mn}=\langle m | \frac{\Lambda}{N} \cdot
                \frac{\partial}{\partial x} | n \rangle.
\end{eqnarray}
We can also derive the equation of motion
\begin{eqnarray}
P_{mn} = \langle m | U'(x) | n \rangle - X_{nm}.
\end{eqnarray} 
In the large $N$ limit, the matrices $X$ and $P$ are replaced with 
the classical functions $X(z, \Lambda)$ and $P(z, \Lambda)$ respectively:
\begin{eqnarray}
P(z, \Lambda) = U'(X(z, \Lambda)) - X(1/z, \Lambda).
\end{eqnarray}

Let us determine the critical potential which realizes the $(4,5)$ model.  
We know that $X$ will be a fourth order differential 
operator in the continuum limit. At the critical point, therefore,
we can set
$X(z,\Lambda=\Lambda_c) = (1-z)^4/z$ and 
$P(z,\Lambda=\Lambda_c)  = \Lambda_c z +$
(higher powers of $z$). 
After substituting these
into the equation of motion, we can find the critical potential
\begin{eqnarray}
U'(\phi) = 8 + 4 \phi + 8 \phi^2 + \phi^3,
\end{eqnarray}
and the critical value of the cosmological constant as $\Lambda_c = 70$.

Off the critical point, we set the classical functions $X(z,\Lambda)$
and $P(z,\Lambda)$ as
\begin{eqnarray}
X(z, \Lambda) & = & \sqrt{R(\Lambda)}/z + a(\Lambda) + b(\Lambda) z
               + c(\Lambda) z^2 + d(\Lambda) z^3,
\nonumber \\
P(z, \Lambda) & = & \Lambda z/\sqrt{R(\Lambda)} + (higher\  powers\ of\ z).
\nonumber
\end{eqnarray}
Here $R(\Lambda)$ is called the specific heat function.
Expanding the equation of motion in powers of $z$,
we obtain the third order equation of $a(\Lambda)$:
\begin{eqnarray}
a(\Lambda)^3 + \ell (R,\Lambda) a(\Lambda)^2 + 
m(R,\Lambda) a(\Lambda)+ n(R,\Lambda) = 0,
\label{eqa3}
\end{eqnarray}
where
\begin{eqnarray}
\ell(R,\Lambda) & = & 8,
\nonumber \\
m(R,\Lambda) & = & 
    \frac{3 + 271 R(\Lambda) + 9 R(\Lambda)^2 - 27 R(\Lambda)^3}
         {1 + 15 R(\Lambda)},
\nonumber \\
n(R,\Lambda) & = & \frac{8 + 40 R(\Lambda) + 24 R(\Lambda)^2 - 72 R(\Lambda)^3}
             {1 + 15 R(\Lambda)}.
\nonumber 
\end{eqnarray}  
Other functions are given in terms of $a(\Lambda)$ as follows
\begin{eqnarray}
b(\Lambda) & = & \frac{\sqrt{R(\Lambda)}}{1-3R(\Lambda)}
                    \left\{
                       4+16a(\Lambda)+3a(\Lambda)^2
                    \right\}, 
\nonumber \\
c(\Lambda) & = & R(\Lambda) \left\{ 8+3a(\Lambda) \right\},
\hspace{1cm}
d(\Lambda)  =  R(\Lambda)^{\frac{3}{2}},
\nonumber \\
\Lambda & = & 3R(\Lambda)^3
            + \left\{
                 18a(\Lambda)^2+96a(\Lambda)+128
              \right\} R(\Lambda)^2
            + \left\{
                 3b(\Lambda)^2-1
              \right\} R(\Lambda)
\nonumber\\
         && + \left\{
                 3a(\Lambda)^2+16a(\Lambda)+4
              \right\}b(\Lambda) \sqrt{R(\Lambda)}.
\nonumber
\end{eqnarray}
The third order equation (\ref{eqa3})
has three possible solutions.
Generally, the $(p,q)$ model has a relation between the
cosmological constant $\Lambda$ and the specific heat $R(\Lambda)$:
\begin{eqnarray}
R(\Lambda) -1 \sim (\Lambda-\Lambda_c)^{2/(p+q-1)}.
\end{eqnarray}   
We should take the solution that satisfies this relation
for the case of the $(p, q) = (4, 5)$ model.

The exact expression of $X(z,\Lambda)$ can determine
the expectation value $ \frac{\Lambda}{N} \langle tr A^n \rangle$.
In the large $N$ limit, the summation is replaced with the
integration: 
\begin{eqnarray}
\frac{\Lambda}{N} \langle tr A^n \rangle
  =  \frac{\Lambda}{N} \sum_{i=1}^N \langle i | x^n | i \rangle
\simeq
\int_0^\Lambda d\lambda \oint_0 \frac{dz}{2 \pi i z} [X(z,\lambda)]^n.
\nonumber \\
\end{eqnarray}


Next, let us consider the three-matrix model.
As in the case of the two-matrix model,
we introduce the orthogonal polynomials $\tilde{\Pi}(x)$ 
which satisfy
\begin{eqnarray}
\langle m | n \rangle & \equiv &
\int dxdydz e^{-S(x,y,z)}
           \tilde{\Pi}(z)  \tilde{\Pi}(x) = \delta_{mn}.
\end{eqnarray}
It is useful to introduce matrices $X_1$, $X_2$, $X_3$ and $P_1$:
\begin{eqnarray}
\left[ X_1 \right]_{mn} & = & \langle m | x | n \rangle ,
\nonumber \\
\left[ X_2 \right]_{mn} & = & \langle m | y | n \rangle 
\hspace{0.4cm} = \hspace{0.4cm} 
\left[ X_2 \right]_{nm} ,
\nonumber \\
\left[ X_3 \right]_{mn} & = & \langle m | z | n \rangle 
\hspace{0.4cm} = \hspace{0.4cm}  
\left[ X_1 \right]_{nm} ,
\nonumber \\
\left[ P_1 \right]_{mn} & = & \langle m | \frac{\Lambda}{N} \cdot
\frac{\partial}{\partial x} | n \rangle .
\nonumber
\end{eqnarray}
With these matrices, the equations of motion are expressed as
\begin{eqnarray}
\left[ P_1 \right]_{mn} & = & \langle m | U'_1(x) | n \rangle -
                              \left[ X_2 \right]_{mn},
\label{3-matrix model ce}\\
\langle m | U'_2(y) | n  \rangle & = & \left[ X_1 \right]_{mn} +
                                     \left[ X_1 \right]_{nm}.  
\nonumber 
\end{eqnarray}
Introducing classical functions,
in the large $N$ limit, 
(\ref{3-matrix model ce}) can be rewritten as
\begin{eqnarray}
P(z, \Lambda) & = & U'_1(X_1(z, \Lambda)) - X_2(z, \Lambda),
\label{eom3mat1}\\
U'_2(X_2(z, \Lambda)) & = & X_1(z, \Lambda) + X_1(1/z, \Lambda).
\label{eom3mat}
\end{eqnarray}
Remark that the classical function satisfies 
$X_2(z,\Lambda) = X_2(1/z,\Lambda)$,
because the matrix $X_2$ is symmetric under the transposition.
Now let us determine the critical potentials
which realize the $(4, 5)$ model.
The classical functions are now
\begin{eqnarray}
X_{1}(z, \Lambda_c) & = & \frac{(1-z/4)(1-z)^4}{z},
\nonumber\\
X_{2}(z, \Lambda_c) & = & - \frac{(1-z)^4}{z^2}.
\nonumber
\end{eqnarray}
By using these critical behaviors and the equations of motion 
(\ref{eom3mat1}), (\ref{eom3mat}),
we obtain the critical potentials and the critical value 
of the cosmological constant,
\begin{eqnarray}
U'_1(\phi) & = &  \frac{111}{16} - \frac{9}{2} \phi - \phi^2,
\nonumber\\
U'_2(\phi) & = &                 - \frac{3}{2} \phi - \frac{1}{4} \phi^2,
\nonumber\\
\Lambda_c & = & 35. 
\end{eqnarray}

The critical potentials determine
the classical functions $X_i(z,\Lambda)$.
We expand them in terms of $z$:
\begin{eqnarray}
X_1(z, \Lambda) & = & \sqrt{R(\Lambda)}/z + \tilde{a}(\Lambda)
                     + \tilde{b}(\Lambda) z + \tilde{c}(\Lambda) z^2 
                     + \tilde{d}(\Lambda) z^3 + \tilde{e}(\Lambda) z^4,
\nonumber\\
X_2(z, \Lambda) & = & \tilde{f}(\Lambda)/{z^2} + \tilde{g}(\Lambda)/z 
                    + \tilde{h}(\Lambda) + \tilde{g}(\Lambda) z 
                    + \tilde{f}(\Lambda) z^2,
\nonumber \\
P(z, \Lambda) & = & \Lambda z/\sqrt{R(\Lambda)} + 
(higher\ powers\ of\ z).
\nonumber
\end{eqnarray}
After substituting these into (\ref{eom3mat1}) and (\ref{eom3mat}),
we get the fourth order equation of
$\tilde{a}(\Lambda)$:
\begin{equation}
\tilde{a}(\Lambda)^4+\tilde{k}(R,\Lambda) \tilde{a}(\Lambda)^3+
\tilde{\ell}(R,\Lambda) \tilde{a}(\Lambda)^2+
\tilde{m}(R,\Lambda) \tilde{a}(\Lambda)+
\tilde{n}(R,\Lambda) = 0,
\end{equation}
where
\begin{eqnarray}
\tilde{k}(R,\Lambda) & = & 
    \frac{288R(\Lambda)^3 + 60R(\Lambda)^2 + 9}{32R(\Lambda)^3+1},
\nonumber\\
\tilde{\ell}(R,\Lambda) & = & 
         \frac{12R(\Lambda)^4 + 972R(\Lambda)^3 + 297R(\Lambda)^2 
                         + 36R(\Lambda) + \frac{8}{3}}
                     {32R(\Lambda)^3+1},
\nonumber\\
\tilde{m}(R,\Lambda) & = & \frac{54R(\Lambda)^4 + 1474R(\Lambda)^3 
                         + \frac{1941}{4}R(\Lambda)^2 
                         + 54R(\Lambda) - \frac{1303}{16}}
                     {32R(\Lambda)^3+1},
\nonumber\\
\tilde{n}(R,\Lambda) & = & 
        \frac{\frac{243}{4}R(\Lambda)^4 + \frac{6849}{8}R(\Lambda)^3 
                         + \frac{4443}{16}R(\Lambda)^2 
                         - \frac{3}{4}R(\Lambda) + \frac{22977}{256}}
                     {32R(\Lambda)^3+1}.
\nonumber
\end{eqnarray}
The solution $\tilde{a}(\Lambda)$ has four possible branches.
As in the case of the two-matrix model,
we require the relation,
$R(z,\Lambda)-1 \sim (\Lambda-\Lambda_c)^{2/(4+5-1)}$,
which determines $\tilde{a}(\Lambda)$ uniquely.  By using
$\tilde{a}(\Lambda)$, the other functions are given as
\begin{eqnarray}
\tilde{b}(\Lambda) & = & 
   \frac{R(\Lambda)^{1/2}
       \left\{
                  1367-12\tilde{a}(\Lambda)-432\tilde{a}(\Lambda)^2
                  -64\tilde{a}(\Lambda)^3
                - \left( 144 + 64 \tilde{a}(\Lambda) \right) R(\Lambda)
       \right\}
         }{32 \left\{ 2+9R(\Lambda)+4\tilde{a}(\Lambda) R(\Lambda) \right\}},
\nonumber \\
\tilde{c}(\Lambda) & = & - \left\{ \frac{3}{2}\tilde{a}(\Lambda)^2
                       + \frac{27}{4}\tilde{a}(\Lambda) + \frac{3}{32}
                           \right\}R(\Lambda)
                         - \tilde{b}(\Lambda) R(\Lambda)^{3/2},
\nonumber\\
\tilde{d}(\Lambda) & = & -\frac{R(\Lambda)^{3/2}}{2}
                            \left\{ 2\tilde{a}(\Lambda) + \frac{9}{2} \right\},
\hspace{0,4cm}
\tilde{e}(\Lambda)  =  -\frac{R(\Lambda)^2}{4},
\hspace{0.4cm}
\tilde{f}(\Lambda)  =  - R(\Lambda),
\nonumber\\
\tilde{g}(\Lambda) & = & - \left\{ 2\tilde{a}(\Lambda) + \frac{9}{2} \right\}
                              R(\Lambda)^{1/2},
\nonumber\\
\tilde{h}(\Lambda) & = & - \tilde{a}(\Lambda)^2 - \frac{9}{2}\tilde{a}(\Lambda)
                      + \frac{111}{16} - 2\tilde{b}(\Lambda) R(\Lambda)^{1/2},
\nonumber\\
\Lambda & = & - \left\{
                   2\tilde{a}(\Lambda) \tilde{b}(\Lambda) 
                  + \frac{9}{2}\tilde{b}(\Lambda)
              \right\} R(\Lambda)^{1/2}
            + \left\{
                   2\tilde{a}(\Lambda)+\frac{9}{2}
              \right\} R(\Lambda)
\nonumber\\
         && + \left\{
                   3\tilde{a}(\Lambda)^2 + \frac{27}{2} \tilde{a}(\Lambda)
                  + \frac{3}{16}
              \right\} R(\Lambda)^2
            + 2\tilde{b}(\Lambda) R(\Lambda)^{5/2}.
\nonumber\\
\end{eqnarray}

The classical functions
$X_1(z,\Lambda)$ and $X_2(z,\Lambda)$
enable us to
evaluate the expectation values 
$\frac{\Lambda}{N}\langle tr A^n \rangle$ 
        ($= \frac{\Lambda}{N}\langle tr C^n \rangle$)
and $\frac{\Lambda}{N}\langle tr B^n \rangle$.
For example, $\frac{\Lambda}{N}\langle tr B^n \rangle$
can be determined by
\begin{eqnarray}
 \frac{\Lambda}{N} \langle B^n \rangle
\simeq   \int_0^\Lambda \oint_0 
                \frac{dz}{2 \pi i z} [X_2(z,\lambda)]^n.
\nonumber
\end{eqnarray}
It is hard to calculate
the expectation values $\frac{\Lambda}{N}\langle tr (A^k B^\ell) \rangle$ and
$\frac{\Lambda}{N}\langle tr (A^k B^\ell C^m) \rangle$
directly.
These expectation values, however, can be reduced to
$\frac{\Lambda}{N}\langle tr A^n \rangle$ 
by using some kinds of the Schwinger-Dyson equations.



\begin{thebibliography}{9}
\bibitem{Pasq}
V.Pasquier, Nucl. Phys. B285 (1987) 162;
J. Phys. 35 (1987) 5707.

\bibitem{HI}
Harish-Chandra, Amer. J. Math. 79 (1957) 87,\\
C. Itzykson and J. -B. Zuber, J. Math. Phys. 21 (1980) 411.

\bibitem{KBK}
V. A. Kazakov, Phys. Lett. A119 (1986) 140,\\
D. Boulatov and V. A. Kazakov, Phys. Lett. B186 (1987) 379.

\bibitem{Douglas}
M. Douglas, ``The Two-matrix Model'',
Proceedings of Cargese Workshop, 1990,\\
T. Tada, Phys. Lett. B259 (1991) 442.

\bibitem{Tada}
T. Tada and M. Yamaguchi, Phys. Lett. B250 (1990) 38.

\bibitem{DKK}
J. -M. Daul, V. A. Kazakov and I. K. Kostov, Nucl. Phys. B409 (1993)
311. 

\bibitem{AIT}
M. Anazawa, A. Ishikawa and H. Tanaka, Prog. Theor. Phys. 98 (1997)
457, hep-th/9705002.

\bibitem{GN}
E. Gava and K. S. Narain, Phys. Lett. B263 (1991) 213.

\bibitem{Staudacher}
M. Staudacher, Phys. Lett. B305 (1993) 332.

\bibitem{IIKMNS}
M. Ikehara, N. Ishibashi, H. Kawai, T. Mogami, R. Nakayama and
N. Sasakura, Phys. Rev. D50 (1994) 7467.

\bibitem{SY}
F. Sugino and T. Yoneya, Phys. Rev. D53 (1996) 4448.

\bibitem{GM}
D. J. Gross and A. A. Migdal, Nucl. Phys. B340 (1990) 333, \\
M. Douglas and S. Shenker, Nucl. Phys. B335 (1990) 635, \\
E. Brezin and V. A. Kazakov, Phys. Lett. B236 (1990) 144.

\bibitem{IK}
N. Ishibashi and H. Kawai, Phys. Lett. B322 (1994) 67.

\end{thebibliography}
\end{document}